\documentclass[nofootinbib,showpacs,preprintnumbers,
amsmath,amssymb,superscriptaddress]{revtex4}

\usepackage{graphicx}

\begin{document}

\title{Electroweak $2 \rightarrow 2$ amplitudes for 
electron-positron annihilation 
at TeV energies}

\vspace*{0.3 cm}

\author{A.~Barroso}
\affiliation{CFTC, University of Lisbon
Av. Prof. Gama Pinto 2, P-1649-003 Lisbon, Portugal}
\author{B.I.~Ermolaev}
\altaffiliation[Permanent Address: ]{Ioffe Physico-Technical Institute, 194021
 St.Petersburg, Russia}
\affiliation{CFTC, University of Lisbon
Av. Prof. Gama Pinto 2, P-1649-003 Lisbon, Portugal}
\author{M.~Greco} 
\affiliation{Dept of Physics and INFN, University Rome III, Rome, Italy} 
\author{S.M.~Oliveira}
\affiliation{CFTC, University of Lisbon
Av. Prof. Gama Pinto 2, P-1649-003 Lisbon, Portugal}
\author{S.I.~Troyan}
\affiliation{St.Petersburg Institute of Nuclear Physics, 
188300 Gatchina, Russia}

\begin{abstract}
The non-radiative scattering amplitudes for electron-positron annihilation 
into quark and lepton pairs in the TeV energy range 
are calculated in the double-logarithmic approximation. The 
expressions for the amplitudes are obtained using infrared evolution 
equations with different cut-offs for virtual photons and for $W$ and $Z$ 
bosons, and compared with previous results obtained with an universal 
cut-off.
\end{abstract}


\maketitle

\section{Introduction}

Next future linear $e^+e^-$ colliders will be operating in a energy 
domain which is much higher than the electroweak bosons masses, so that 
the full knowledge of the scattering amplitudes for $e^+e^-$
annihilation into quark and lepton pairs will be needed. 
The forward-backward asymmetry for $e^+e^-$ 
annihilation into leptons or
hadrons produced at energies much greater than the $W$ and $Z$ boson 
masses 
 has been recently considered in Ref.~\cite{egt}, 
where the electroweak radiative corrections 
were calculated to all orders in the double-logarithmic approximation (DLA). 
It was shown that the effect of the electroweak 
DL radiative corrections on the value of the 
forward-backward asymmetry is quite sizable and grows rapidly 
with the energy. 
As usual, the asymmetry is defined as the 
difference between the forward and the backward scattering amplitudes
over the sum of them. 
These amplitudes were calculated in Ref.~\cite{egt} in DLA, 
 by introducing and solving the Infrared Evolution Equations 
(IREE). 
 This method is a very simple and the most efficient instrument 
for performing all-orders 
double-logarithmic calculations (see Ref.~\cite{flmm} and
Refs. therein). In particular, when it was  
applied in Ref.~\cite{flmm} to 
calculate the electroweak Sudakov (infrared-divergent) logarithms, 
 it led easily to the proof of the exponentiation of the Sudakov 
logarithms. 
 At that moment this was in 
contradiction to the non-exponentiation claimed in Ref.~\cite{pciaf} and 
obtained 
by other means. This contradiction 
provoked a large discussion about the exponentiation. 
The exponentiation was confirmed eventually 
by the two-loop calculations in Refs.~\cite{me}-\cite{dp} and 
by summing up the higher loop DL contributions in Refs.~\cite{kp} and \cite{m}.
These Sudakov logarithms provide the whole set of DL 
contributions to the $2 \to 2$ amplitudes only 
when the process is considered in the 
hard kinematic region where all the 
Mandelstam variables $s,t,u$ are of the same order. 
On the other hand, when the kinematics of the $2 \to 2$ processes 
is of the Regge type, besides the Sudakov logarithms, 
another kind of DL contributions arises, coming from ladder Feynman 
graphs. 
Accounting for those (infrared stable) contributions it leads, instead of 
simple exponentials, to much 
more complicated expressions for the scattering amplitudes. 
This was first shown in Ref.~\cite{ggfl}, where in the framework of 
pure QED, the scattering amplitudes for the forward and backward 
$e^+e^- \to \mu^+\mu^-$ annihilation were calculated
in the Regge kinematics. 
One example of high-energy 
electroweak processes in the Regge kinematics 
was considered in Ref.~\cite{flmm}, where the backward scattering 
amplitude was calculated, for the annihilation 
of a lepton pair with same helicities into another pair of leptons.
 More general calculations of the 
forward and backward electroweak scattering amplitudes were done in 
Ref~\cite{egt}. 

However, both calculations in 
Refs.~\cite{egt} and ~\cite{flmm} were done under the assumption 
 that the transverse momenta $k_{i \perp}$ of the virtual photons and 
virtual $W,Z$ -bosons were much greater than 
the masses of the weak bosons. In 
other words, the same infrared cut-off $M$ 
in the transverse momentum space, 
 was used for all virtual electroweak bosons, i.e., 
\begin{equation}
\label{1cut}
k_{i\perp} \gg M \geq M_W \approx M_Z .
\end{equation}

 Obviously, while $M$ is the natural 
infrared cut-off for the logarithmic contributions involving $W,Z$ 
bosons, the cut-off for the photons can be chosen independently. in 
accord with the experimental resolution in a given observed process. 
Indeed the assumption (\ref{1cut}), although simplifying the calculations 
a lot, is 
unnecessary and an approach that involves different cut-offs for 
photons and $W,Z$ 
weak bosons would be more interesting and suitable for phenomenological 
applications. This technique 
involving different cut-offs for photons and for $W,Z$ bosons was applied
in Ref.~\cite{flmm}, for calculating the double-logarithmic 
contributions of soft 
virtual electroweak bosons (the Sudakov electroweak logarithms) but 
not for the scattering amplitudes in the regions of Regge kinematics. 
 
In the present paper we generalize the results of Refs.~\cite{flmm} and 
~\cite{egt}, and obtain new double-logarithmic expressions for 
the $2 \to 2$ - 
electroweak amplitudes in the 
forward and backward kinematics. These expressions involve 
therefore different 
infrared cut-offs for virtual photons 
and virtual weak 
bosons. 
Throughout the paper we assume that the photon cut-off, $\mu$, and the 
$W,Z$ boson cut-off, $M$, satisfy the relations 
\begin{equation}
\label{cuts}
M \geq M_{W,Z},~~~\mu \geq m_f
\end{equation}
where $m_f$ is the largest mass of the quarks or leptons involved in the 
process. 
Notice that the 
values of $M$ and $\mu$ could be widely different.
Let us remind that in order to study a scattering amplitude $A(s,t)$ in 
the Regge kinematics $s \gg -t$ (where $s$ and $t$ are the standard 
Mandelstam variables), it is convenient to represent $A(s,t)$ in 
the following form: $A(s,t) = A^{(+)}(s,t)+ A^{(-)}(s,t)$, 
with $ A^{(\pm)}(s,t) = (1/2)[A(s,t) \pm A(-s,t)]$ called the positive 
(negative) signature amplitudes. We shall consider only amplitudes
with the positive signatures. The IREE for the 
negative signature electroweak amplitudes can be obtained in a similar way, 
see e.g. Ref.~\cite{egt} for more details. 

The paper is organized as follows: in Sect.~2 we define the kinematics 
 and express the scattering amplitude for the $e^+e^-$ annihilation 
in terms of invariant amplitudes. 
In Sect.~3, we construct the evolution equations for the invariant 
amplitudes for 
the case when in the center mass (cm) frame, the scattering angles are 
very small. 
First, we obtain the IREE 
equations in the integral form and then we transform them in 
the simpler, differential form. 
These differential equations are solved in Sect.~4 and explicit expressions 
 for the invariant amplitudes involving the Mellin integrals are 
obtained. 
In Sect.~5, we consider the case of large scattering angles, or when the 
Mandelstam variables s, t and u are all large. 
Sect.~6 deals with the expansion of 
the invariant amplitudes into the perturbative series in order to 
extract the first-loop and the second-loop contributions. Then we compare 
these contributions to 
the analogous terms obtained when one universal cut-off is used and 
study their difference. 
The effect of high-order contributions in the two 
approaches is further studied 
in Sect.~7 where the asymptotic expressions of the amplitudes are compared. 
Finally, Sect.~8 contains our concluding remarks.

\section{Invariant amplitudes for the annihilation processes}
\label{ELAST}

Let us 
consider a general process 
where the lepton $l^k(p_1)$ and its anti-particle $\bar{l}_i(p_2)$ 
annihilate into a quark or a lepton 
$q ^{k'}(p'_1)$ and its anti-particle $\bar{q}_{i'}({p'}_2)$ (see Fig.~1): 
\begin{equation}
\label{annih}
l^k(p_1)\bar{l}_i(p_2) \to q^{k'}({p'}_1) \bar{q}_{i'}({p'}_2) ~. 
\end{equation} 

\begin{figure}[htbp]
 \begin{center}
 \includegraphics{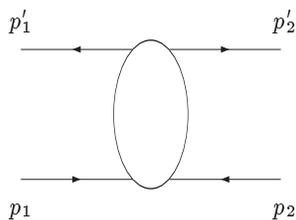}
 \caption{Scattering amplitude of the annihilation of Eq. 3.}
 \label{fig.T1}
 \end{center}
\end{figure}

For this process, the most complicate case occurs 
when both the initial and the final particles (anti-particles) 
are left-handed (right-handed). 
The scattering amplitudes for other helicities can be obtained easily from 
the formulae derived for this case. 
As there is no technical difference when considering 
the annihilation into 
quarks or leptons, we present parallel results for the 
annihilation into a quark-antiquark or a 
lepton-antilepton pair. According to our assumption, the initial lepton 
belongs to the weak isodoublet $(\nu,~e)$. The final lepton belongs to 
another doublet, e.g. $(\nu_{\mu},~\mu)$, and the final quarks are also from 
a doublet, e.g. $(u, ~d)$. The antilepton and the antiquark belong to 
the charge conjugate doublets. 
Obviously, the scattering amplitude $A$ for the 
annihilation can be written as follows:
\begin{equation}
\label{A}
A = q^{k'}({p'}_1) \bar{q}_{i'}({p'}_2) A^{i i'}_{k'
  k}l^k(p_1)\bar{l}_i(p_2) ~ ,
\end{equation}
where the $SU(2)$ matrix 
amplitude $A^{i i'}_{k' k}$ has to be calculated. We will 
consider it in DLA. The 
DL contributions to $A^{i i'}_{k' k}$ are different according to the 
kinematics of the process. The kinematics is defined by appropriate 
relations among the Mandelstam variables $s,~t,~u$, 
\begin{equation}
\label{studef}
s = (p_1 + p_2)^2,~t = (p_1 - {p'}_1)^2, ~u = (p_1 - {p'}_2)^2 ~.
\end{equation}
Throughout this paper we assume that $\sqrt{s} \gg M_{W,Z}$. 

The kinematical regime defined as 
\begin{equation}
\label{hard}
-t \sim -u \sim s
\end{equation} 
is called the hard kinematics and corresponds to large cm scattering angles 
$\theta \equiv \theta_{ \mathbf{p_1\,{p'}_1}} \sim 1$ . Radiative corrections 
to the annihilation in this kinematics yield DL contributions. 
 
There are also two other Regge-type kinematical regimes where DL 
contributions appear. 
First, there is the configuration where 
\begin{equation}
\label{tkin}
 s \sim -u \gg -t ~.
\end{equation} 
We call it the $t$ -kinematics. According to the terminology introduced in 
 \cite{ggfl}, it is the forward 
kinematics (with respect to the charge flow) for 
$e^+e^- \to \mu^+ \mu^-$ and $e^+e^- \to d \bar{d}$. At the same time, it 
corresponds to the backward kinematics for $e^+e^- \to u \bar{u}$. 
In this kinematics, $\theta \ll 1$. 

Second, there is the opposite kinematics where $\theta \sim\pi$ and therefore 
\begin{equation}
\label{ukin}
 s \sim -t \gg -u ~.
\end{equation}

We define the configuration (\ref{ukin}) as the 
$u$-kinematics. It corresponds to the 
forward scattering for $e^+e^- \to u \bar{u}$ and the backward scattering 
for $e^+e^- \to \mu^+ \mu^-$,~ $e^+e^- \to d \bar{d}$.
 
To simplify the calculations, it is convenient to introduce the 
projection operators $(P_j)^{i i'}_{k' k} ~(j = 1,2,3,4)$, so that 
 $A^{i i'}_{k' k}$ can be written in the following form:
\begin{equation}
\label{invampl}
A^{i i'}_{k' k} = \frac{\bar{u}(-{p'}_2) \gamma_{\mu} u({p'}_1)
\bar{u}(-p_2) \gamma^{\mu} u(p_1)}{s}
|\Big[ (P_j)_{kk'}^{ii'} A_j + (P_{j + 1})_{kk'}^{ii'} A_{j + 1}\Big] ~,
\end{equation}
where $j = 1$ for the $t$-kinematics and $j = 3$ for the $u$-kinematics. 
The representation (\ref{invampl}) reduces the calculation of the 
matrix amplitude $A^{i i'}_{k' k}$ to the calculation of the invariant 
amplitudes $A_j$. 
The explicit expressions for the operators $(P_c)^{i i'}_{k' k} ~(c = 1,..,4)$ 
can be taken from Ref.~\cite{egt}:
\begin{eqnarray}
\label{pr}
(P_1)_{kk'}^{ii'} = \frac12\delta_k^{i'} \delta^i_{k'},
 &&(P_2)_{kk'}^{ii'} = 
2 (t_c)_k^{i}(t_c)^{i'}_{k'}, \\ \nonumber
(P_3)_{kk'}^{ii'} = \frac{1}{2} \left[\delta^i_k \delta^{i'}_{k'}-
\delta^{i'}_{k} \delta^i_{k'}\right]~,
 &&(P_4)_{kk'}^{ii'} = \frac{1}{2} \left[\delta^i_k \delta_{k'}^{i'} +
\delta^{i'}_{k} \delta^i_{k'}\right] ~.
\end{eqnarray}

According to the results of Ref.~\cite{egt}, the forward $(A_F)$ and backward 
$(A_B)$ amplitudes of the 
$e^+e^-$ annihilation into quarks are expressed through invariant 
amplitudes $A_j$ as follows:
\begin{eqnarray}
\label{amplquark}
A_F(e^+e^- \to u\bar{u}) = (A_3 + A_4)/2,
~A_B(e^+e^- \to u\bar{u}) = A_4 , \\ \nonumber
~A_F(e^+e^- \to d\bar{d}) = (A_1 + A_2)/2, 
~A_B(e^+e^- \to d\bar{d}) = A_2~.
\end{eqnarray} 
and the annihilation into leptons is expressed through the leptonic invariant 
amplitudes very similarly: 
\begin{eqnarray}
\label{ampllept}
A_F(e^+e^- \to \mu^+\mu^-) = (A_1 + A_2)/2, 
~A_B(e^+e^- \to \mu^+\mu^-) = A_2 ~, \\ \nonumber
A_F(e^+e^- \to \nu_{\mu} \bar{\nu_{\mu}}) = (A_3 + A_4)/2,
~~A_B(e^+e^- \to \nu_{\mu} \bar{\nu_{\mu}}) = A_4 ~.
\end{eqnarray} 
We have used the general notation $A_j$ for the invariant amplitudes in 
Eqs.~(\ref{amplquark}, \ref{ampllept}) and we will keep using this 
notation until Sect.~4. 

In order to calculate the amplitudes $A_j$ to all orders in the
electroweak couplings in the DLA, we construct and solve some infrared
evolution equations (IREE). 
 These equations describe the evolution of $A_j,~(j = 1,2,3,4)$ with 
respect to an infrared cut-off. We introduce two such cut-offs, $\mu$ 
and $M$. We presume that $M \approx M_Z \approx M_W$ and use this 
cut-off to regulate the DL contributions involving soft 
(almost on-shell) virtual $W,Z$ -bosons. 
In order to regulate the IR divergences arising from soft photons we use 
the cut-off $\mu$ and we assume that 
$\mu \approx m_q \ll M$ where $m_q$ is the maximal quark mass involved. 
Both cut-offs are introduced in the transverse momentum space (with 
respect to the plane formed by momenta of the initial leptons) so that 
the transverse momenta $k_i$ of virtual photons obey
\begin{equation}
\label{mu}
k_{i\perp} > \mu ~,
\end{equation}
while the momenta $k_i$ of virtual $W,Z$ -bosons obey 
\begin{equation}
\label{M}
k_{i\perp} > M ~.
\end{equation}

Let us first consider $A_j$ in the collinear 
 kinematics where, in the cm frame, the produced quarks or leptons move 
very 
close to the $e^+e^-$ -beams. 
In order to fix such kinematics, we implement Eq.~(\ref{tkin}) 
by the further restriction on $t$: 
\begin{equation}
\label{tmu}
s \sim -u \gg M^2 \gg \mu^2 \geq -t
\end{equation}
and similarly for Eq.~(\ref{ukin}) by 
\begin{equation}
\label{umu}
s \sim -t \gg M^2 \gg \mu^2 \geq -u ~. 
\end{equation}

Basically in DLA, the invariant amplitudes $A_j$ 
depend on 
$s, u$ and $t$ through logarithms. Under the 
restriction imposed by Eqs.~(\ref{tmu}, \ref{umu}) then all $A_j$
depend only on logarithms of $s, M^2, \mu^2$ in the collinear kinematics. 
It is convenient to represent $A_j$ in the following form: 
\begin{equation}
\label{ampl}
A_j(s, \mu^2, M^2) = A_j^{(QED)}(s, \mu^2) 
+ {A'}_j(s, \mu^2, M^2) ~,
\end{equation} 
where $A_j^{(QED)}(s, \mu^2)$ accounts for QED DL contributions only, 
i.e. the contributions of 
Feynman graphs without virtual $W,Z$ bosons. To calculate 
$A_j^{(QED)}(s, \mu^2)$ we use the cut-off $\mu$, therefore the amplitudes 
$A_j^{(QED)}$ do not depend on $M$. 
In contrast, the 
amplitudes ${A'}_j(s, \mu^2, M^2)$ depend on both cut-offs. These 
amplitudes account for 
 DL contributions of the Feynman graphs, with one or more $W,Z$ propagators. 
By technical reasons, 
it is convenient to introduce two auxiliary 
amplitudes. The first one, $\tilde{A}^{(QED)}_j(s, M^2)$, is the same QED 
amplitude but with a cut-off $M$. 
The second auxiliary amplitude, 
$\tilde{A}_j(s, M^2)$ accounts for all electroweak DL contributions and
the cut-off $M$ is used to regulate both the virtual photons and 
the weak bosons infrared divergences. 
 
Beyond the Born approximation, the invariant amplitudes we have 
introduced depend on logarithms, 
 the arguments of which can be chosen as in 
the following parameterization:  
\begin{eqnarray}
\label{param}
A_j^{(QED)} = A_j^{(QED)}(s, \mu^2) = A_j^{(QED)}(s/ \mu^2), 
~\tilde{A}^{(QED)}_j = \tilde{A}^{(QED)}_j(s, M^2) = 
\tilde{A}^{(QED)}_j(s/ M^2), \\ \nonumber
~\tilde{A}_j = \tilde{A}_j(s, M^2) = \tilde{A}_j(s/ M^2), 
 ~{A'}_j = {A'}_j (s, \mu^2, M^2) = {A'}_j (s/M^2,\eta) ~, 
\end{eqnarray}
with 
\begin{equation}
\label{eta}
\eta \equiv \ln(M^2/\mu^2) ~. 
\end{equation}

Our aim is to calculate the amplitudes ${A'}_j$, whereas the amplitudes 
$A_j^{(QED)}, \tilde{A}^{(QED)}_j$ and $\tilde{A}_j(s/ M^2)$ are 
supposed to be known. 
The amplitudes 
$\tilde{A}_j$ were introduced and calculated in Ref.~\cite{egt}. 
In order to define amplitudes 
 ${A'}_j, \tilde{A}_j(s, M^2)$, the projection 
operators of Eq.~(\ref{pr}) have been used. The use of these operators is 
based on
 the fact that the $SU(2)\times U(1)$ symmetry for the electroweak 
scattering amplitudes takes place at energies much higher than the 
weak mass scale $M$. On the contrary, the QED amplitudes $A_j^{(QED)}$ and 
$\tilde{A}^{(QED)}_j$ are not $SU(2)$ invariant at any energy.
Nevertheless, it is convenient to introduce 
``the QED invariant amplitudes''
$A_j^{(QED)}, \tilde{A}^{(QED)}_j$ by explicit calculation of 
the forward and backward QED 
scattering amplitudes. Then inverting 
Eq.~(\ref{amplquark}), we construct the amplitudes $A_j^{(QED)}$ for
$e^+e^-$- annihilation into quarks:
\begin{eqnarray}
\label{qedquark}
A_1^{(QED)} = 
2 A_F^{(QED)}(e^+e¯- \to d \bar{d}) - A_B^{(QED)}(e^+e¯- \to u \bar{u})
,
~~A_2^{(QED)} = A_B^{(QED)}(e^+e¯- \to u \bar{u}) , \\ \nonumber 
A_3^{(QED)} = 
2A_F^{(QED)}(e^+e¯- \to u \bar{u}) - A_B^{(QED)}(e^+e¯- \to d \bar{d})
,
~~A_4^{(QED)} = A_B^{(QED)}(e^+e¯- \to d \bar{d}) ~ 
\end{eqnarray} 
 and inverting Eq.~(\ref{ampllept}) allows us to obtain $A_j^{(QED)}$ for
 $e^+e^-$- annihilation into leptons:
\begin{eqnarray}
\label{qedlept}
A_1^{(QED)} = 
2A_F^{(QED)}(e^+e¯- \to \mu^+ \mu^-), 
~A_2^{(QED)} = 0, 
~A_3^{(QED)} = - A_4^{(QED)} = A_B^{(QED)}(e^+e¯- \to \mu^+ \mu^-).
\end{eqnarray} 

\section{Evolution equations for amplitudes $A_j$ in the collinear 
kinematics}
 
We would like to discuss now the IREE for the amplitudes introduced 
earlier.
The basic idea for constructing infrared evolution equations for 
the scattering amplitudes consists in introducing 
the infrared cut-offs in the transverse momentum space and evolving the 
scattering amplitudes with respect to them. 
This method does not involve analyzing the DL contributions 
of specific Feynman graphs but is based on quite general
conceptions such as the analyticity of the scattering amplitudes and 
the dispersion relations which guarantees its applicability to a wide 
class of problems (see e.g. Ref.~\cite{flmm} and Refs. therein). 
The essence of the method is the factorizing the DL contributions of virtual 
particles with the minimal transverse momenta. 
The IREE with two cut-offs for the electroweak amplitudes in the hard
kinematics (\ref{hard}) were obtained in Ref.~\cite{flmm}. In the present section
we construct the IREE for the $2\rightarrow 2$ - electroweak
amplitudes in the Regge kinematics.
According to 
Eqs.~(\ref{mu}, \ref{M}), we use two different cut-offs for the virtual photons 
and for the weak bosons. 
The amplitude $A_j$ is in the lhs of such an equation. The rhs 
contains several terms. In the first place, there is the 
Born amplitude $B_j$. In order to obtain the other terms in 
the rhs, we use the fact that the DL contributions of virtual particles 
with minimal transverse 
momenta $( \equiv k_{\perp})$ 
can be factorized. 
Furthermore, this $k_{\perp}$ acts as a
new cut-off for the other virtual momenta. The virtual 
particle with $k_{\perp}$ (we call such a particle the softest one) can 
be either an electroweak bosons or a fermion. 
Let us suppose first that the softest particle is an electroweak boson. 

In this case, 
in the Feynman gauge, DL contributions come from the graphs where 
the softest 
propagator is attached to the external lines in every possible way 
whereas $k_{\perp}$ acts as a new cut-off for the blobs as shown 
in Fig.~2. 
\begin{figure}[htbp]
 \begin{center}
 \includegraphics{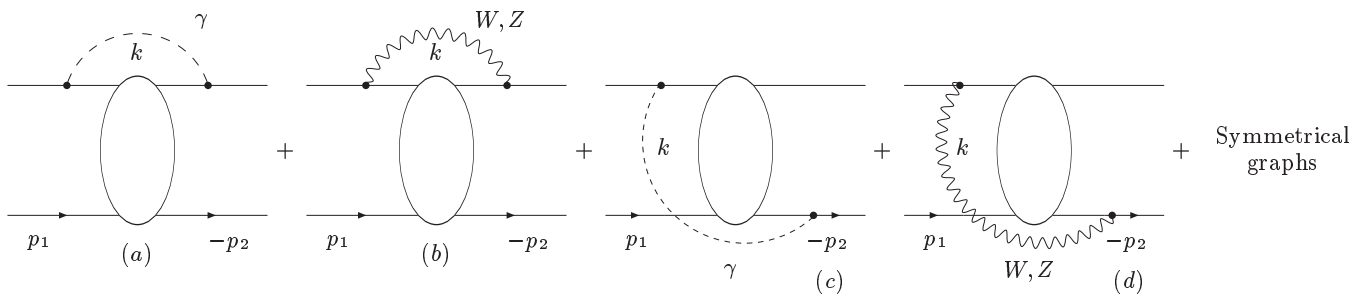}
 \caption{Softest boson contributions to IREE to $A_j$.}
 \end{center}
\end{figure}
When the softest electroweak boson is a photon, the integration region 
over $k_{\perp}$ is 
$\mu^2 \ll k_{\perp}^2 \ll s$ 
 and its contribution, $G^{\gamma}_j$, to the rhs of the IREE is: 
\begin{eqnarray}
\label{gphot}
G^{(\gamma)}_j = - \frac{1}{8\pi^2} b_j^{(\gamma)} 
&\Big(&\int_{\mu^2}^s \frac{dk_{\perp}^2 }{k_{\perp}^2} 
\ln(s/k_{\perp}^2)A^{(QED)}_j(s, k_{\perp}^2) +
\int_{\mu^2}^{M^2} \frac{dk_{\perp}^2 }{k_{\perp}^2}\ln(s/k_{\perp}^2)
{A'}_j(s, k_{\perp}^2, M^2) + \\ \nonumber 
&&\int_{M^2}^s \frac{dk_{\perp}^2 }{k_{\perp}^2} 
\ln(s/k_{\perp}^2) A'_j(s, k_{\perp}^2, k_{\perp}^2 )\, \Big) ~,
\end{eqnarray} 
where 
\begin{eqnarray}
\label{bphot}
b_1^{(\gamma)} = g^2 \sin^2 \theta_W \frac{(Y_2 - Y_1)^2}{4}, 
~~~b_2^{(\gamma)} = g^2 \sin^2 \theta_W \Big[ \frac{1}{6} + 
\frac{(Y_2 - Y_1)^2}{4} \Big] ~,\\ \nonumber
b_3^{(\gamma)} = g^2 \sin^2 \theta_W \frac{(Y_2 + Y_1)^2}{4}, 
~~~b_4^{(\gamma)} = g^2\sin^2 \theta_W \Big[ \frac{1}{6} + 
\frac{(Y_2 + Y_1)^2}{4} \Big] ~. \\ \nonumber
\end{eqnarray}

We have used the standard notations in Eq.~(\ref{bphot}): $g,~ g'$ are
the Standard Model couplings, $Y_1 ~(Y_2)$ is the hypercharge of
the initial (final) fermions and $\theta_W$ is the Weinberg angle.
The 
logarithmic factors in the integrands of Eq.~(\ref{gphot}) correspond to 
the integration in the longitudinal momentum space. 
The amplitude $A'$ in the last integral of Eq.~(\ref{gphot}) does not
depend on $\mu$ because $k_{\perp}^2 > M^2$. Therefore it can be
expressed in terms of $\tilde{A}_j(s,k_{\perp}^2)$ and
$\tilde{A}_j^{(QED)}(s,k_{\perp}^2)$ :
\begin{equation}
  {A'}_j(s, k_{\perp}^2, k_{\perp}^2) = \tilde{A}_j(s,
  k_{\perp}^2)-\tilde{A}_j^{(QED)}(s,k_{\perp}^2)~.
\end{equation}

When the softest boson is either a $Z$ or a $W$, 
its DL contribution 
can be factorized in the region $M^2 \ll k_{\perp}^2 \ll s$. This 
yields: 
\begin{equation}
\label{gwz}
G_j^{(WZ)} = - \frac{1}{8\pi^2} b_j^{(WZ)}
\int_{M^2}^s \frac{dk_{\perp}^2 }{k_{\perp}^2} 
\ln(s/k_{\perp}^2)\tilde{A}_j(s/k_{\perp}^2 ) ~, 
\end{equation} 
with 
\begin{equation}
\label{bwz}
b_j^{(WZ)} = b_j - b_j^{\gamma} ~ 
\end{equation}
and the factors $b_j$ can be taken from Ref.~\cite{egt}: 
\begin{eqnarray}
\label{btot}
b_1 = \frac{{g'}^2 (Y_1 - Y_2)^2}{4}, 
~~b_2 = \frac{8g^2+{g'}^2 (Y_1 - Y_2)^2}{4}, \\ \nonumber
~~b_3=\frac{{g'}^2 (Y_1 + Y_2)^2}{4},
~~b_4 = \frac{8g^2 + {g'}^2 (Y_1 + Y_2)^2}{4}~~.
\end{eqnarray}

In Eq.~(\ref{gwz}) we have used the fact that the $W$ and the 
$Z$ bosons cannot 
be the softest particles for the amplitudes $A^{(QED)}_j$ 
since the integrations over the softest transverse momenta 
in $A^{(QED)}_j$ can go down to $\mu$, by definition. 
The sum of Eqs. (\ref{gphot}) and (\ref{gwz})
, $G_j$ can be written in the 
more convenient way: 
\begin{equation}
\label{gtot}
G_j(s, \mu^2, M^2) = 
G^{(\gamma)}_j(s, \mu^2, M^2)+ G_j^{(WZ)}(s,M^2) =
G_j^{(QED)}(s, \mu^2) - 
\tilde{G}^{(QED)}_j(s, M^2) + \tilde{G}_j(s, M^2) + {G'}_j(s, \mu^2, M^2) ~,
\end{equation} 
where 
\begin{eqnarray}
\label{g}
G^{(QED)}_j = - \frac{1}{8\pi^2} b_j^{(\gamma)} 
\int_{\mu^2}^s \frac{dk_{\perp}^2 }{k_{\perp}^2} 
\ln(s/k_{\perp}^2)A^{(QED)}_j(s/k_{\perp}^2) ~,
\tilde{G}^{(QED)}_j = -\frac{1}{8\pi^2} b_j^{(\gamma)} 
\int_{M^2}^s \frac{dk_{\perp}^2 }{k_{\perp}^2} 
\ln(s/k_{\perp}^2)\tilde{A}^{(QED)}_j (s/k_{\perp}^2) ~, \\ \nonumber 
\tilde{G}_j = - \frac{1}{8\pi^2} b_j \int_{M^2}^s
\frac{dk_{\perp}^2 }{k_{\perp}^2} 
\ln(s/k_{\perp}^2)\tilde{A}_j (s/k_{\perp}^2) ~,
{G'}_j = -\frac{1}{8\pi^2} b_j^{(\gamma)} 
\int_{\mu^2}^{M^2} \frac{dk_{\perp}^2 }{k_{\perp}^2} 
\ln(s/k_{\perp}^2){A'}_j (s/k_{\perp}^2, M^2/k_{\perp}^2 ) ~. 
\end{eqnarray}

Eqs.~(\ref{gtot}, \ref{g}) account for DL contributions 
 when the softest particle is an electroweak boson. However, the softest 
particle can also be a virtual fermion. In this case, DL contributions 
from the integration over the 
momentum $k$ of the softest fermion arise 
from the diagram shown in Fig.~3 where the amplitudes 
$A_j$ are factorized into two on-shell amplitudes in the $t$-channel. 
We denote this contribution by $Q_j(s, \mu^2, M^2)$.
\begin{figure}[htbp]
 \begin{center}
 \includegraphics{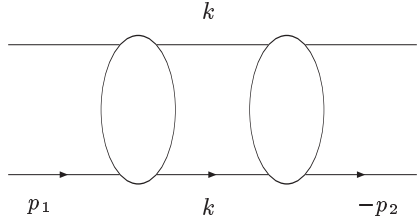}
 \caption{Softest fermion contribution.}
 \end{center}
\end{figure}

The analytic expression for $Q_j$ is rather 
cumbersome. However it looks simpler when the Sudakov parameterization 
is introduced for the softest quark momentum $k$
(with $p_1$ and $p_2$ being the initial lepton momenta).
\begin{equation}
\label{sudak}
k = \alpha p_2 + \beta p_1 + k_{\perp}~.
\end{equation} 
After simplifying the spin structure, we obtain 
\begin{equation}
\label{q}
Q_j(s, \mu^2, M^2) = 
c_j \int_{\mu^2}^s 
d k_{\perp}^2 \int \frac{d \alpha}{\alpha}\frac{ d \beta}{\beta} 
\frac{2k_{\perp}^2 }{(s\alpha\beta - k_{\perp}^2)^2} 
A_j(s\alpha, k_{\perp}^2, M^2)
A_j(s\beta, k_{\perp}^2, M^2) ~, 
\end{equation}
where 
\begin{equation}
\label{c}
c_1 = c_2 = -c_3 = -c_4 = \frac{1}{8 \pi^2}~.
\end{equation}

Similarly to Eq.~(\ref{gtot}), $Q_j$ of 
 Eq.~(\ref{q}) can be divided into the following simple contributions:
\begin{equation}
\label{qtot}
Q_j = Q_j^{(QED)} - \tilde{Q}_j^{(QED)} + \tilde{Q}_j + {Q'}_j ~,
\end{equation}
where 
\begin{equation}
\label{qqed}
Q_j^{(QED)} (s/\mu^2)= 
c_j \int_{\mu^2}^s 
d k_{\perp}^2 \int \frac{d \alpha}{\alpha}\frac{ d \beta}{\beta} 
\frac{k_{\perp}^2 }{(s\alpha\beta - k_{\perp}^2)^2} 
A_j^{(QED)}(s\alpha/ k_{\perp}^2)
A_j^{(QED)}(s\beta/ k_{\perp}^2) ~, 
\end{equation}

\begin{equation}
\label{qqedtilde}
 \tilde{Q}_j^{(QED)}(s/ M^2) = 
c_j \int_{M^2}^s 
d k_{\perp}^2 \int \frac{d \alpha}{\alpha}\frac{ d \beta}{\beta} 
\frac{k_{\perp}^2 }{(s\alpha\beta - k_{\perp}^2)^2} 
 \tilde{A}_j^{(QED)}(s\alpha/ k_{\perp}^2)
 \tilde{A}_j^{(QED)}(s\beta/ k_{\perp}^2) ~, 
\end{equation}

\begin{equation}
\label{qtilde}
 \tilde{Q}_j(s/ M^2) = 
c_j \int_{M^2}^s 
d k_{\perp}^2 \int \frac{d \alpha}{\alpha}\frac{ d \beta}{\beta} 
\frac{k_{\perp}^2 }{(s\alpha\beta - k_{\perp}^2)^2} 
 \tilde{A}_j(s\alpha/ k_{\perp}^2)
 \tilde{A}_j(s\beta/ k_{\perp}^2) ~, 
\end{equation}
and 
\begin{eqnarray}
\label{qprime} 
 {Q'}_j(s/M^2, \eta) = 
c_j \int_{\mu^2}^{M^2} 
d k_{\perp}^2 \int \frac{d \alpha}{\alpha}\frac{ d \beta}{\beta} 
\frac{2k_{\perp}^2 }{(s\alpha\beta - k_{\perp}^2)^2} 
\Big(2 A_j^{(QED)}(s\alpha/ k_{\perp}^2 )
 {A'}_j(s\beta/ k_{\perp}^2, M^2/k_{\perp}^2) \\ \nonumber 
 + {A'}_j(s\alpha/ k_{\perp}^2, M^2/k_{\perp}^2) 
 {A'}_j(s\beta/ k_{\perp}^2, M^2/k_{\perp}^2) \Big)~. 
\end{eqnarray}

Now we are able to write the IREE for amplitudes $A_j$. The general form 
it given by:
\begin{equation}
\label{ireegeneral}
A_j = B_j + G_j + Q_j ~.
\end{equation}
 
Then using Eqs.~(\ref{gtot}) and (\ref{qtot}) we can rewrite it as 
\begin{equation}
\label{iree}
{A'}_j + A_j^{(QED)}= B_j^{(QED)} - \tilde{B}_j^{(QED)} + \tilde{B}_j 
+ G_j^{(QED)} - \tilde{G}_j^{(QED)} + \tilde{G}_j + {G'}_j
+ Q_j^{(QED)} - \tilde{Q}_j^{(QED)} + \tilde{Q}_j + {Q'}_j ~.
\end{equation}

Let us notice that $A_j^{(QED)}(s/\mu^2)$ obeys the equation 
\begin{equation}
\label{ireeqed}
A_j^{(QED)} = B_j^{(QED)} + G_j^{(QED)} + Q_j^{(QED)}
\end{equation}
and therefore $A_j^{(QED)}$ cancels out in Eq.~(\ref{iree}). 
Also, the auxiliary amplitudes $\tilde{A}_j$ and 
$\tilde{A}_j^{(QED)}$, 
obey similar equations: 
\begin{equation}
\label{ireem}
\tilde{A}_j^{(QED)} = \tilde{B}_j^{(QED)} + \tilde{G}_j^{(QED)} + 
\tilde{G}_j^{(QED)}, ~~
\tilde{A}_j = \tilde{B}_j + \tilde{G}_j + \tilde{Q}_j ~.
\end{equation}

The solutions to Eqs.~(\ref{ireeqed}, \ref{ireem}) are known. 
With the notations that we have used 
they can be taken directly from Ref.~\cite{egt}. 
Hence, we are left with 
the only unknown 
amplitude ${A'}_j$ in Eq.~(\ref{iree}). Using 
Eqs.~(\ref{ireeqed}, \ref{ireem}), we arrive at the IREE for ${A'}_j$, 
namely: 
\begin{equation}
\label{ireeaprime}
{A'}_j(s/M^2, \eta) = \tilde{A}_j(s/M^2) - \tilde{A}_j^{(QED)}(s/M^2) + 
{G'}_j(s/M^2, \eta) + {Q'}_j(s/M^2, \eta) ~. 
\end{equation}
 
In order to solve Eq.~(\ref{ireeaprime}), it is more convenient to 
use the Sommerfeld-Watson 
transform. As long as one considers the positive signature amplitudes, 
this 
transform formally coincides with the Mellin transform. 
It is convenient to 
use different forms of this transform for 
the invariant amplitudes we consider: 
\begin{eqnarray}
\label{mellinaqed}
A_j^{(QED)}(s/ \mu^2) &=& 
\int_{-\imath \infty}^{\imath \infty} \frac{d \omega}{2\pi\imath}\,
\Big( \frac {s}{\mu^2}\Big)^{\omega} f_j^{(0)} (\omega) ~, 
\\ \nonumber \\
\label{mellinaqedtilde}
\tilde{A}_j^{(QED)}(s/ M^2) &=& 
\int_{-\imath \infty}^{\imath \infty} \frac{d \omega}{2\pi\imath}\,
\Big( \frac {s}{M^2}\Big)^{\omega} f_j^{(0)}(\omega) ~,
\\ \nonumber \\
\label{mellinatilde}
 \tilde{A}_j(s/ M^2) &=& 
\int_{-\imath \infty}^{\imath \infty} \frac{d \omega}{2\pi\imath}\,
\Big( \frac {s}{M^2}\Big)^{\omega} f_j(\omega) ~,
\\ \nonumber \\
\label{mellinaprime}
{A'}_j(s/M^2, \eta) &=& 
\int_{-\imath \infty}^{\imath \infty} \frac{d \omega}{2\pi\imath}\,
\Big( \frac {s}{M^2}\Big)^{\omega} F_j (\omega, \eta) ~. 
\end{eqnarray} 

Combining Eqs.~(\ref{mellinaqed}) to (\ref{mellinaprime}) 
with Eq.~(\ref{ireeaprime})
we arrive at the following equation for the Mellin amplitude 
$F_j (\omega, \eta)$: 
\begin{eqnarray}
\label{masterint}
\int_{-\imath \infty}^{\imath \infty} \frac{d \omega}{2\pi\imath}\,
\Big( \frac {s}{M^2}\Big)^{\omega} 
F_j (\omega, \eta) &=& 
\int_{-\imath \infty}^{\imath \infty} \frac{d \omega}{2\pi\imath}\,
\Big( \frac {s}{M^2}\Big)^{\omega} 
[f_j(\omega) - f_j^{(0)}(\omega)] \\ \nonumber
& & -\int_{-\imath \infty}^{\imath \infty} \frac{d \omega}{2\pi\imath}\,
\Big( \frac {s}{M^2}\Big)^{\omega} \frac{1}{8\pi^2} b_j^{(\gamma)} 
\int_{\mu^2}^{M^2} \frac{dk_{\perp}^2 }{k_{\perp}^2} 
\ln(s/k_{\perp}^2)F_j (\omega, \eta' ) \\ \nonumber
& &+\int_{-\imath \infty}^{\imath \infty} \frac{d \omega}{2\pi\imath}\,
\Big( \frac {s}{M^2}\Big)^{\omega} c_j\int_{\mu^2}^{M^2} \frac{dk_{\perp}^2 }{k_{\perp}^2} 
\Big( 2f_j^{(0)}(\omega)F_j (\omega, \eta' ) + 
F_j^2(\omega, \eta') \Big) \Big] ~,
\end{eqnarray}
where $\eta' = \ln(M^2/k_{\perp}^2)$. 
Differentiating Eq.~(\ref{masterint}) with respect to $\mu^2$ leads to the 
homogeneous partial differential equation for the on-shell amplitude 
$F_j (\omega, \eta)$: 
\begin{equation}
\label{master}
\frac{\partial F_j}{\partial \eta} = 
 - \frac{1}{8\pi^2} b_j^{(\gamma)} 
\Big( - \frac{\partial F_j}{\partial \omega} + \eta F_j\Big) + 
c_j \Big( 2 f_j^{(0)}(\omega) F_j + F_j^2 \Big)~,
\end{equation}
where we have used the fact that $\ln(s/\mu^2)$, 
in Eq.~(\ref{masterint}), can be rewritten as 
 $\ln(s/M^2) + \eta$ and that $\ln(s/M^2)$ corresponds to 
$- \partial/ \partial \omega$~. 

\section{Solutions to the evolution equations for collinear kinematics}

Let us consider first the particular case when 
$b_1^{(\gamma)} = 0$. It contributes to the forward leptonic, 
$e^+e^- \to \mu^+\mu^-$ annihilation and corresponds, in our notations, to 
the option 
\begin{equation}
\label{leptons}
 ~~Y_1 = Y_2 = -1.
\end{equation} 
Let us notice that $A_j$ with $j = 1$ contributes also to the forward 
$e^+e^- \to d \bar{d}$ annihilation, though here $Y_1 = -1, Y_2 = 1/3$ 
and therefore $b_1^{(\gamma)} \neq 0$. 
In order to avoid confusion between these cases, we change our notations, 
denoting $\Phi_1 \equiv F_1,~ \phi_1 \equiv f_1$ and 
$\phi_1^{(0)} \equiv f_1^{(0)}$ when $Y_1 = Y_2 = -1$. We will also use 
notations $\Phi_{2,3,4}$ instead of $F_{2,3,4}$ when we discuss the 
annihilation into 
leptons. Then we denote 
$c \equiv c_1 = 1/(8 \pi^2)$. 
Therefore, the lepton amplitude $\Phi_1 (\omega, \eta)$ 
for the particular case 
 (\ref{leptons}) obeys the Riccati equation 
\begin{equation}
\label{eqf1}
\frac{\partial \Phi_1}{\partial \eta} = 
c \Big( 2 \phi_1^{(0)}(\omega) \Phi_1 + \Phi_1^2 \Big)~,
\end{equation}
with the general solution 
\begin{equation}
\label{f1general}
\Phi_1 = 
\frac{e^{2c \phi_1^{(0)}\eta}}{C \phi_1^{(0)} - 
e^{2c \phi_1^{(0)}\eta}/2 \phi_1^{(0)} } ~. 
\end{equation} 
In order to specify $C$, we use the matching (see Eq.~(\ref{masterint}))
\begin{equation}
\label{match}
\Phi_1 = \phi_1(\omega) - \phi_1^{(0)}(\omega)~,
\end{equation}
when $\eta = 0$, arriving immediately at 
\begin{equation}
\label{Phi1}
\Phi_1 = 
\frac{2\phi_1^{(0)}(\phi_1 -\phi _1^{(0)})e^{2c\phi_1^{(0)}\eta}}
{\phi_1^{(0)} + \phi_1 - (\phi_1 -\phi_1^{(0)})e^{2c\phi_1^{(0)}\eta}} 
\end{equation}
and therefore to the following expression for the invariant amplitude 
$L_1 \equiv A_1$ when $Y_1 = Y_2 = -1$: 
\begin{equation}
\label{aleptf}
L_1 = 
\int_{-\imath \infty}^{\imath \infty} \frac{d \omega}{2\pi\imath}\,
\Big( \frac {s}{\mu^2}\Big)^{\omega} \phi_1^{(0)}(\omega) + 
\int_{-\imath \infty}^{\imath \infty} \frac{d \omega}{2\pi\imath}\,
\Big( \frac {s}{M^2}\Big)^{\omega}
\frac{2\phi_1^{(0)}(\phi_1 -\phi _1^{(0)})e^{2c\phi_1^{(0)}\eta}}
{\phi_1^{(0)} + \phi_1 - (\phi_1 -\phi_1^{(0)})e^{2c\phi_1^{(0)}\eta}} ~.
\end{equation} 
Obviously, when $\mu \to M$, Eqs.~(\ref{aleptf}) converges to the same 
amplitude obtained with using only one cut-off. Indeed, substituting 
$\mu = M$ and $\eta = 0$ leads to 
\begin{equation}
\label{aleptf1}
L_1 = 
\int_{-\imath \infty}^{\imath \infty} \frac{d \omega}{2\pi\imath}\,
\Big( \frac {s}{M^2}\Big)^{\omega} \phi_1(\omega)~.
\end{equation}

According to Eqs.~(\ref{qedlept}, \ref{mellinaqed}), 
the QED amplitude $\phi_1^{(0)}$ is easily expressed in terms of 
 Mellin amplitude 
$\phi_F^{(0)}$ for the forward 
$e^+e^- \to \mu^+\mu^-$ annihilation: 
\begin{equation}
\label{phi0f}
\phi_1^{(0)} = 2 \phi_F^{(0)}~.
\end{equation}
The expression for 
$\phi_F^{(0)}$ 
can be taken from Refs.~\cite{ggfl},\cite{flmm} and \cite{egt}: 
\begin{eqnarray}
\label{phi0}
\phi_F^{(0)} = 
4\pi^2 \big( \omega - \sqrt{\omega^2 - \chi^2_0} \big), 
\end{eqnarray}
with 
\begin{equation}
\label{chi0}
\chi^2_0 = 2\alpha/\pi.
\end{equation} 
On the other hand, the amplitudes $\phi_j$ were calculated in 
Ref.~\cite{egt}. In particular, 
\begin{equation}
\label{phi1}
\phi_1 = 4 \pi^2\big( \omega - \sqrt{\omega^2 - \chi^2} \big)~,
\end{equation}
where $\chi^2$ is expressed through the electroweak couplings $g$ and 
$g'$: 
\begin{equation}
\label{chi}
\chi^2 = [3 g^2 + {g'}^2]/(8 \pi^2) ~. 
\end{equation} 

Next, let us solve Eq.~(\ref{master}) for the general case of non-zero 
factor $b_j^{(\gamma)}$. Then, this equation describes 
the backward $e^+ e^-$ annihilation 
into a lepton pair (e.g. $\mu^+ \mu^-$) and also the forward and 
backward annihilation into quarks. 
 Eq.~(\ref{master}) looks simpler when $\omega$, $\eta$ are replaced by 
new variables 
\begin{equation}
\label{xy}
x = \omega/\lambda_j,~~~y = \lambda_j \eta~,
\end{equation}
with $\lambda_j = \sqrt{ b_j^{(\gamma)}/(8\pi^2)}$. 
Changing to the new 
variables, we arrive again at the Riccati equation: 
\begin{equation}
\label{riccatij}
\frac{\partial F_j}{\partial \tau} = 
\big( \sigma - \tau \big) F_j -2q_j f^{(0)}_j F_j - q_j F^2_j ~,
\end{equation}
where $\sigma = (x + y)/2,~~ \tau = (x - y)/2$ and 
$q_j = c_j /\lambda_j$.

The general solution to Eqs.~(\ref{riccatij}) is 
\begin{equation}
\label{generalj}
F_j = \frac{P_j(\sigma, \tau)}{ C(\sigma) + q_j Q_j(\sigma, \tau)}~,
\end{equation}
where $C(\sigma)$ should be specified, 
\begin{equation}
\label{P}
P_j(\sigma, \tau) = \exp \Big( \sigma\tau - \tau^2/2 
 - 2q_j \int_{\sigma}^{\sigma + \tau} d \zeta f_j^{(0)}(\zeta) \Big) 
\end{equation} 
and 
\begin{equation}
\label{Q}
Q_j (\sigma, \tau) = 
\int_{\sigma}^{\sigma + \tau} d \zeta P_j(\sigma, \zeta) ~.
\end{equation}

The QED amplitudes $f^{(0)}_j$ can be obtained from the known expressions 
for the backward, $f^{(0)}_B$ and forward $f^{(0)}_j$ QED scattering 
amplitudes: 
\begin{eqnarray}
\label{fzerob}
 f^{(0)}_B(x) = (4\pi\alpha e_q/ p_B^{(0)})
d \ln (e^{x^2/4} D_{p_B^{(0)}}
(x)) /d x ~, 
\end{eqnarray} 
where $D_p$ are the Parabolic cylinder functions 
with $ p_B^{(0)} = -2e_q/(1 + e_q)^2$ and $e_q = 1$ for the 
annihilation into muons, $e_q = 1/3 ~(2/3)$ for the annihilation into 
$d ~(u)$- quarks. Similarly, the QED forward scattering amplitudes for the 
annihilation into quarks are 
\begin{eqnarray}
\label{fzerof}
 f^{(0)}_F(x) = (4\pi\alpha e_q / p_F^{(0)})
d \ln (e^{x^2/4} D_{p_F^{(0)}}(x))/d x ~, 
\end{eqnarray} 
with $ p_B^{(0)} = 2e_q/(1 - e_q)^2$ .
Let us stress that the forward amplitudes for the annihilation into leptons 
are given by Eq.~(\ref{aleptf}). The
amplitude $f^{(0)}_{F,B}$ was obtained first in Ref.~\cite{ggfl} for the 
backward scattering in QED. 
Obviously, the only difference between the formulae 
for $f_j(x)$ and $f^{(0)}_j(x)$ is 
in the different factors $a_j$, $p_j $ and $\lambda_j$. 
We can specify $C(\sigma)$, using the matching
\begin{equation}
\label{matchj}
F_j(\omega) = f_j(\omega) - f^{(0)}_j(\omega) ~,
\end{equation}
when $\eta = 0$. The invariant amplitudes $f_j$ were calculated in 
Ref.~\cite{egt}:
\begin{equation}
\label{fj}
f_j (x) = \frac{a_j}{ p_j}
\frac{d \ln (e^{x^2/4} D_{p_j}(x)) }{d x} 
= a_j \frac{ D_{p_j - 1}(x))}{D_{p_j}(x))} ~.
\end{equation}

Using Eq.~(\ref{matchj}) we are led to 
\begin{equation}
\label{Fj}
F_j = \frac{\big( f_j(x + y) - f^{(0)}_j (x + y)\big) P(\sigma, \tau)}
{P(\sigma, \sigma) - (f_j(x + y) - f^{(0)}_j (x + y)) 
\big( Q(\sigma, \sigma) - Q(\sigma, \tau) \big)} ~ 
\end{equation} 
and finally to 
\begin{eqnarray}
\label{solutionaj}
A_j(s/M^2, \eta) &=& 
\int_{-\imath \infty}^{\imath \infty} \frac{d \omega}{2\pi\imath}\,
\Big( \frac {s}{\mu^2}\Big)^{\omega} f_j^{(0)} (\omega) + 
\\ \nonumber
& &\int_{-\imath \infty}^{\imath \infty} \frac{d \omega}{2\pi\imath}\,
\Big( \frac {s}{M^2}\Big)^{\omega}
\frac{\big( f_j(x + y) - f^{(0)}_j (x + y)\big) P_j(\sigma, \tau)}
{P_j(\sigma, \sigma) - (f_j(x + y) - f^{(0)}_j (x + y)) 
\big( Q_j(\sigma, \sigma) - Q_j(\sigma, \tau) \big)} ~. 
\end{eqnarray}
It is easy to check that when $\mu = M$, 
$A_j(s/M^2, \eta)$ coincides with the amplitude $\tilde{A}_j(s, M^2)$ 
obtained in Ref.~\cite{egt}. 

Eqs.~(\ref{aleptf}, \ref{solutionaj}) describe all invariant amplitudes 
for $e^+e^-$ -annihilation into a quark or a lepton pair in the collinear 
kinematics~(\ref{tmu}, \ref{umu}). 
 
\section{Scattering amplitudes at large values of $t$ and $u$ }
\label{FB}

In this section we calculate the scattering amplitudes 
$A$ when the 
restriction of Eqs.~(\ref{tmu}, \ref{umu}) for the 
kinematical configurations (\ref{tkin}, \ref{ukin}) are replaced by 
\begin{equation}
\label{tm}
s \gg M^2 \geq -t \gg \mu^2 
\end{equation}
and 
\begin{equation}
\label{um}
s \gg M^2 \geq -u \gg \mu^2 ~. 
\end{equation} 

In this kinematical regions
it is more convenient to study the 
scattering amplitudes $A$ directly, rather than using 
the invariant amplitudes $A_j$. 
In order to unify the discussion for 
 both kinematics~(\ref{tm}, \ref{um}), let us introduce 
\begin{equation}
\label{kappat}
\kappa = -t ~,
\end{equation} 
when (\ref{tm}) is considered 
and
\begin{equation}
\label{kappau}
\kappa = -u 
\end{equation}
for the other case (\ref{um}). 
Using this notation, the same parameterization 
$A = A(s, \mu^2, M^2, \kappa)$ 
holds for both kinematics~(\ref{tm}, \ref{um}). 
Let us discuss now the evolution equations for $A$. 
As in the previous case, it is convenient 
to consider separately the purely 
QED part, $A^{(QED)}$ and the mixed part, $A'$: 
\begin{equation}
\label{separ}
A(s, \kappa, \mu^2, M^2) = A^{(QED)}(s, \kappa, \mu^2) 
+ A'(s, \kappa, \mu^2, M^2 ). 
\end{equation}

Generalizing Eq.~(\ref{param}), we can parameterize them as follows:
\begin{equation}
A^{(QED)}(s, \kappa, \mu^2) = A^{(QED)}(s/\mu^2, \kappa/\mu^2),~~~~~~~ 
A'(s, \kappa, \mu^2, M^2 ) = A'(s/M^2, s/\mu^2, \kappa/ \mu^2, M^2/\mu^2 ). 
\end{equation}
 
In order to construct the 
IREE for $A^{(QED)}$ and $A'$, we should consider again all options for the softest 
virtual particles. 
The Born terms for the configurations~(\ref{tm}) and (\ref{um}) do 
not 
depend on $\mu^2$ and vanish after differentiating on $\mu$. The same is 
true for the softest quark contributions. Indeed, the softest 
fermion pair yields DL contributions in the integration region 
$k_{\perp}^2 \gg \kappa$, which is unrelated to $\mu$. 
Hence, we are left with the only option for the softest 
particle to be an electroweak boson. 
The factorization region for this kinematics is 
\begin{equation}
\label{kappak}
\mu^2 \ll k_{\perp}^2 \ll \kappa ~.
\end{equation}

Obviously, only virtual photons can be factorized in this factorization 
region, which leads to a simple IREE: 
\begin{eqnarray}
\label{ireekappa}
\frac{\partial A^{(QED)}}{ \partial
\rho} + \frac{\partial A^{(QED)}}{ \partial z} = 
- \lambda ({b^{(\gamma)} \rho+ h^{(\gamma)} z}) A^{(QED)},~~~~
\frac{\partial A'}{ \partial
\rho} + \frac{\partial A'}{ \partial z} + 
\frac{\partial A'}{ \partial \eta'} = 
- \lambda ({b^{(\gamma)} \rho+ h^{(\gamma)} z}) A' 
\end{eqnarray}
where we have denoted $\rho = \ln(s/\mu^2)$, $z = \ln(\kappa/ \mu^2)$,
$ \eta' = \ln(\eta) = \ln(M^2/\mu^2)$ and $\lambda=\alpha/2\pi$. The
factors $b^{(\gamma)}$ and $h^{(\gamma)}$ are: 
\begin{equation}
\label{bht}
h^{(\gamma)} = e_1{e'}_1 + e_2{e'}_2, 
~~b^{(\gamma)} = e_1 e_2 +{e'}_1{e'}_2 - e_2{e'}_1 - e_1{e'}_2 ~
\end{equation}
for the case (\ref{kappat}), and
\begin{equation}
\label{bhu}
h^{(\gamma)} = 
- e_2{e'}_1 + e_1{e'}_2, 
~~b^{(\gamma)} = 
e_1 e_2 +{e'}_1{e'}_2 +e_2{e'}_2 + e_1{e'}_1 ~
\end{equation}
for the other case (\ref{kappau}). 
 
The notations $e_i, {e'}_i$ in Eqs.~(\ref{bht}, \ref{bhu}) 
stand for the absolute values of the electric charges. They correspond to the 
notations of the external particle momenta introduced in Fig.~1. 
The terms proportional to $h^{(\gamma)}$ in Eq.~(\ref{ireekappa}) correspond 
to the Feynman graphs where the 
softest photons propagate in the $\kappa$-channels. Let us notice that 
for any kinematics we consider it holds 
\begin{equation}
\label{bhsum}
b_j^{(\gamma)}+ h_j^{(\gamma)} = (1/2)[e_1^2 + e_2^2 + {e'}_1^2 +{e'}_2^2] 
\end{equation}
due to the electric charge conservation. 

In order to solve Eq.~(\ref{ireekappa}), we use the matching with 
the amplitude 
 $\hat{A} (s, \mu^2, M^2)$ for the same process, however in the collinear 
kinematics: 
\begin{equation}
\label{matchkappa}
A^{(QED)}(s, \mu^2, \kappa, M^2) = \hat{A}^{(QED)}(s,\mu^2),~~ 
A' (s, \kappa, \mu^2, M^2) = \hat{A}'(s, \mu^2, M^2)~,
\end{equation}
when $ \kappa = \mu^2$. The solution to Eq.~(\ref{ireekappa}) is 
\begin{equation}
\label{akappageneral}
A^{(QED)} = \psi^{(QED)}(\rho - z) 
e^{- \lambda b_j^{(\gamma)} \rho^2/2 - \lambda h_j^{(\gamma)} z^2/2},~~
A' = \psi'(\rho - z, \eta' -z) 
e^{-\lambda b_j^{(\gamma)} \rho^2/2 - \lambda h_j^{(\gamma)} z^2/2} .
\end{equation}
Using the matching of Eq.~(\ref{matchkappa}) allows to specify 
$\psi$ and $\psi^{(QED)}$. After that we obtain:
\begin{equation}
\label{akappaprime}
 A^{(QED)} = S' \hat{A}^{(QED)}(s/\kappa) ,~~~~ 
A' = S' \hat{A}'(s/M^2, M^2/\kappa) ~,
\end{equation}
where 
\begin{equation}
\label{sprime}
S' = e^{ -\lambda b_j^{(\gamma)}\rho z +\lambda ( b_j^{(\gamma)}- h_j^{(\gamma)}) z^2/2} ~.
\end{equation}
We did not change $s/M^2$ to $s/\kappa$ in Eq.~(\ref{akappaprime}) 
because $M^2 \gg \kappa$. 
It is convenient to absorb the term $-\lambda b_j^{(\gamma)}\rho z$ into 
the amplitudes $\hat{A}^{(QED)}$ and $\hat{A}'$. Introducing, instead of $\omega$, 
the new Mellin variable $l = \omega + \lambda b_j^{(\gamma)} z$ 
(see Ref.~\cite{egt} for details), we rewrite 
 Eq.~(\ref{akappaprime}) as follows (for the sake of simplicity we 
keep the same notations for these new amplitudes 
$\hat{A}^{(QED)}$ and $\hat{A}'$): 
\begin{equation}
\label{akappa}
 A^{(QED)} = S \big( \hat{A}^{(QED)}(s/\kappa) + 
\hat{A}'(s, \kappa, \mu^2, M^2) \big)
\end{equation}
with $S$ being the Sudakov form factor for the case under discussion. 
$S$ includes the softest, infrared divergent DL contributions. 
When the photon infrared cut-off $\mu$ is assumed to be greater than the 
masses 
of the involved fermions, this form factor is: 
\begin{equation}
\label{sudffbigmu}
S = \exp\Big(- \frac{\lambda}{2} 
\big(b^{(\gamma)}+ h^{(\gamma)}\big) \ln^2(\kappa/\mu^2)
\Big) ~.
\end{equation}

However, in the case of $e^+e^-$ annihilation into quarks (muons), 
if the cut-off $\mu$ is chosen to be very small, less 
than 
the electron mass, $m_e$ the exponent in Eq.~(\ref{sudffbigmu}) 
should be changed to : 
\begin{equation}
\label{sudff}
S = \exp\Big(- \frac{\lambda}{2} 
\big(b^{(\gamma)}+ h^{(\gamma)}\big) 
(\ln^2(\kappa/\mu^2) - \ln^2(m^2_e/\mu^2) - \ln^2(m^2/\mu^2)
\Big) ~,
\end{equation}
where $m$ is the mass of the produced quark or lepton
(cf. Ref.~\cite{Greco:1980mh}).

If $m > \mu >m_e$, the last term in the exponent of Eq.~(\ref{sudff})
is absent. 
The kinematics with larger values of $\kappa$, e.g. 
$s \gg \kappa \gg M^2$, can be studied similarly, although it is 
more convenient to use the 
invariant amplitudes $\hat{A}_j$. 
The result is 
\begin{equation}
\label{akappam}
 \hat{A}_j = S_j\tilde{A}_j(s/\kappa) ~,
\end{equation}
where 
\begin{equation} 
\label{sudffm}
S_j = 
\exp\Big[
-\frac{\lambda}{2} \Big(
\big(b_j^{(\gamma)}+ h_j^{(\gamma)}\big) 
(\ln^2(\kappa/\mu^2) - \ln^2(m^2_e/\mu^2) - \ln^2(m^2/\mu^2)
+\big(b_j - b_j^{(\gamma)} + h_j - h_j^{(\gamma)}\big) \ln^2(\kappa/M^2) 
\Big)\Big] ~ 
\end{equation}
and $\tilde{A}(s/M^2)$ is the scattering amplitude of the same process in 
the limit of collinear kinematics and using a single cut-off $M$. These 
amplitudes were 
defined in Sect.~2. 
The factors $h_j$ given below were calculated in Ref~\cite{egt}: 
\begin{eqnarray}
\label{h}
h_1 = g^2(3 + \tan^2 \theta_W Y_1 Y_2)/2, 
&& h_2 = g^2 (-1 + \tan^2 \theta_W Y_1 Y_2)/2 ~, \\ \nonumber 
h_3 = g^2(3 - \tan^2 \theta_W Y_1 Y_2)/2,
&& h_4 = g^2(-1 - \tan^2 \theta_W Y_1 Y_2)/2 ~.
\end{eqnarray}

The form factors $S$, $S_j$ include
the soft DL contributions, with the cm energies of virtual particles 
ranging from $\mu^2$ to $\kappa$. Due to gauge invariance, the 
sums 
$b_j^{(\gamma)}+ h_j^{(\gamma)}$ 
and 
$b_j + h_j$ do not depend 
on $j$ and $S_j$ is 
actually the same for every invariant amplitude contributing 
to $A^{i i'}_{k' k}$ in the forward (backward) kinematics 
(see Ref.~\cite{egt}). 
Obviously, in the case of the hard kinematics where (see Eq.~(\ref{hard})) 
$s \sim -u \sim -t$, i.e. $s \sim \kappa$, 
ladder graphs do not yield DL contributions. The easiest way to see this, 
is to notice that the factor $(s/\kappa)^{\omega}$ in the 
the Mellin integrals~(\ref{mellinatilde}) 
for amplitudes $\tilde{A}_j$ does not depend on $s$ in 
the hard kinematics, therefore all Mellin integrals do not depend on $s$. 
So, the only source of DL terms in this kinematics 
is given by the 
Sudakov form factor $S_j$ given by Eq.~(\ref{sudffm}). 
 Therefore, we easily arrive at 
the known result 
\begin{equation}
\label{ahardm}
A^{i i'}_{k' k} = B^{i i'}_{k' k} S_j~. 
\end{equation} 
$B^{i i'}_{k' k}$ in 
 Eq.~(\ref{ahardm}) stands for the Born terms. 
The electroweak Sudakov 
form factor~(\ref{sudffm}) with two infrared cut-offs 
 was obtained in 
Ref.~\cite{flmm}. 

\section{Forward $e^+e^-$ annihilation into leptons}
 
Eqs.~(\ref{amplquark}, \ref{ampllept}, \ref{aleptf}) and (\ref{solutionaj}) 
give the explicit expressions for 
the scattering amplitudes of $e^+ e^-$-annihilation into quarks and leptons 
in the collinear kinematics. 
These expressions 
resume the DL contributions to all orders in the electroweak 
couplings and operate with two infrared cut-offs. 
In order to estimate the impact of the two-cuts approach, 
we compare these results to 
the formulae for the same scattering amplitudes obtained in 
Ref.~\cite{egt} where one universal cut-off $M$ was used. 
We focus on the particular case of 
 the scattering amplitudes for 
the forward 
$e^+e^-$ annihilation into leptons and restrict 
ourselves, for the sake of simplicity, 
to the collinear 
kinematics of Eq.~(\ref{tmu}). Other 
amplitudes, and other kinematics can be considered in a very similar way. 
Eqs.~(\ref{ampllept}, \ref{aleptf}) and (\ref{solutionaj}) show
that the scattering amplitude $L_F^{(\mu)}$ of 
the forward $e^+e^-$ into $\mu^-\mu^+$ is 
\begin{eqnarray}
\label{amplmu}
 L_F^{(\mu)}&=& 
\int_{-\imath \infty}^{\imath \infty} \frac{d \omega}{2\pi\imath}\,
\Big( \frac {s}{\mu^2}\Big)^{\omega} \phi_F^{(0)}(\omega) + 
\frac{1}{2}
\int_{-\imath \infty}^{\imath \infty} \frac{d \omega}{2\pi\imath}\,
\Big( \frac {s}{M^2}\Big)^{\omega}
\frac{4\phi_F^{(0)}(\phi_1 -2\phi _F^{(0)})e^{4c\phi_F^{(0)}\eta}}
{2\phi_F^{(0)} + \phi_1 - (\phi_1 -2\phi_F^{(0)})e^{4c\phi_F^{(0)}\eta}} 
\\ \nonumber 
&&+ \frac{1}{2}
\int_{-\imath \infty}^{\imath \infty} \frac{d \omega}{2\pi\imath}\,
\Big( \frac {s}{M^2}\Big)^{\omega}
\frac{\phi_2(x + y) P_2(\sigma, \tau)}
{P_2(\sigma, \sigma) - \phi_2(x + y)\big[Q_2(\sigma, \sigma) - 
Q_2(\sigma, \tau)\big]} ~.
\end{eqnarray} 

The first integral in this equation accounts for purely 
 QED double-logarithmic contributions 
and depends on the QED cut-off 
$\mu$ 
whereas the next integrals sum up mixed QED and weak double-logarithmic 
terms and depend on both $\mu$ and $M$. 
The first and the second integrals in Eq.~(\ref{amplmu}) 
grow with $s$ whilst the 
last integral rapidly falls when $s$ increases. The point 
is that this term actually is
the amplitude for the backward annihilation into muon neutrinos. 
It is easy to check that the QED amplitudes 
$\phi_F^{(0)}$ vanish 
when $\mu = M$ 
and the total integrand contains only 
$[\phi_1(\omega) + \phi_2 (\omega)]/2$. 
In contrast to Eq.~(\ref{amplmu}), purely QED contributions are absent 
in formulae for $e^+e^- $ annihilation into neutrinos. For example, 
the scattering amplitude $L_F^{(\nu)}$ of the 
forward $e^+e^- \to \nu_{\mu}\bar{\nu_{\mu}}$ -annihilation in the 
collinear kinematics is 
\begin{eqnarray}
\label{amplnu}
L_F^{(\nu)} = \frac{1}{2}
\int_{-\imath \infty}^{\imath \infty} \frac{d \omega}{2\pi\imath}\,
\Big( \frac {s}{M^2}\Big)^{\omega}
&\Big[& \frac{\phi_3 (x + y)P_3(\sigma, \tau)}
{P_3(\sigma, \sigma) - \phi_3(x + y) 
[Q_3(\sigma, \sigma) - Q_3(\sigma, \tau)]} + \\ \nonumber 
&&\frac{\phi_4 (x + y)P_4(\sigma, \tau)}
{P_4(\sigma, \sigma) - \phi_4(x + y) 
[Q_4(\sigma, \sigma) - Q_4(\sigma, \tau)]} ~\Big] ~.
\end{eqnarray}
 
Similarly to Eq.~(\ref{amplmu}), the integrand in 
Eq.~(\ref{amplnu}) is equal to 
$[\phi_3(\omega) + \phi_4 (\omega)]/2$ when $\mu = M$. 
Although formally Eqs.~(\ref{amplmu}, \ref{amplnu}) 
correspond to the exclusive 
 $e^+e^-$ annihilation into two leptons, actually these expressions 
also describe the inclusive processes when the emission of photons with 
cm energies $< \mu$ is accounted for. 
 
Let us study the impact of our two-cut-offs approach on 
the scattering amplitude $L_F^{(\mu)}$ of Eq.~(\ref{amplmu}). 
As the last integral in 
Eq.~(\ref{amplmu}) rapidly falls with $s$, it is neglected in our 
estimates and we consider contributions of the first and the second 
integrals only. 
First we compare 
the one-loop and two-loop contributions. 
Such contributions can be easily obtained expanding the rhs of 
Eq.~(\ref{amplmu}) into a perturbative series. 
From Eqs.~(\ref{phi0}) and (\ref{phi1}) one obtains that 
\begin{eqnarray} 
\label{phi01series}
\phi_F^{(0)} &\approx& 2\pi^2 
\Big( \frac{\chi_0^2}{\omega} + 
 \frac{1}{4}\frac{\chi_0^4}{\omega^3} + 
 \frac{1}{8}\frac{\chi_0^6}{\omega^5} + ... \Big) ~, \\ \nonumber
\phi_1 &\approx& 2\pi^2 
\Big( \frac{\chi^2}{\omega} + 
 \frac{1}{4}\frac{\chi^4}{\omega^3} + 
 \frac{1}{8}\frac{\chi^6}{\omega^5} + ... \Big) ~,
\end{eqnarray} 
with $\chi_0$, $\chi$ defined in Eqs.~(\ref{chi0}, \ref{chi}). 
Substituting these series into the first and the second integrals of 
Eq.~(\ref{amplmu}) and performing the 
integrations over $\omega$, we arrive at 
\begin{equation}
\label{afirstloop}
 L^{(1)} = 
\gamma_1^{(1)}\ln^2(s/\mu^2) + 
\gamma_2^{(1)} \ln(s/\mu^2)\ln(s/M^2) + 
\gamma_3^{(1)}\ln^2(s/M^2) 
\end{equation} 
for the first-loop contribution to $L_F^{(\mu)}$ and 
\begin{eqnarray}
\label{asecondloop}
L^{(2)} &=& 
\gamma_1^{(2)} \ln^4(s/\mu^2) + 
\gamma_2^{(2)} \ln^3(s/\mu^2)\ln(s/M^2) + \\ \nonumber 
&&\gamma_3^{(2)} \ln^2(s/\mu^2)\ln^2(s/M^2) + 
\gamma_4^{(2)} \ln(s/\mu^2)\ln^3(s/M^2) + 
\gamma_5^{(2)} \ln^4(s/ M^2) 
\end{eqnarray} 
for the second-loop contribution.
The coefficients $\gamma_i^{(k)}$ are given below: 
\begin{eqnarray}
\label{gammaik}
\gamma_1^{(1)} = {\frac{{{\pi }^2}\,{{{{\chi}_0}}^4}}{4}},\, 
\gamma_2^{(1)} = {\frac{{{\pi }^2}\,
 \left( {{\chi}^4} - 4\,{{{{\chi}_0}}^4} \right) }{4}},\, 
\gamma_3^{(1)} = - {\frac{{{\pi }^2}\,
 \left( {{\chi}^4} - 6\,{{{{\chi}_0}}^4} \right) }{8}} ,\, \\ \nonumber 
\gamma_1^{(2)} = {\frac{{{\pi }^2}\,{{{{\chi}_0}}^6}}{96}}, \, 
\gamma_2^{(2)} = 0, \,
\gamma_3^{(2)} = {\frac{{{\pi }^2}\,{{\chi}^2}\,
 \left( {{\chi}^4} - 4\,{{{{\chi}_0}}^4} \right) }{32}},\, \\ \nonumber
\gamma_4^{(2)} = -{\frac{{{\pi }^2}\,\left( {{\chi}^6} - 
 6\,{{\chi}^2}\,{{{{\chi}_0}}^4} + 2\,{{{{\chi}_0}}^6} \right)
 }{24}}, \,
 \gamma_5^{(2)} = {\frac{{{\pi }^2}\,
 \left( 3\,{{\chi}^6} - 24\,{{\chi}^2}\,{{{{\chi}_0}}^4} + 
 14\,{{{{\chi}_0}}^6} \right) }{192}}.
\end{eqnarray} 

Let us compare the above results with those obtained with one universal 
cut-off $M$ only. We 
introduce the notation $\tilde{L}(s/M^2)$ for amplitude $L_F^{(\mu)}$ 
when one cut-off $M$ is used. 
The ratio $R^{1} = L^{1}(s,\mu,M)/\tilde{L}^{(1)}(s,M) $ of the first loop 
contributions to the amplitudes $L_F^{(\mu)}$ and $\tilde{L}$ is 

\begin{equation}
\label{R1}
 R^{(1)} = \frac{L^{(1)}}
{\tilde{\gamma}^{1} \ln^2(s/M^2)} ~ 
\end{equation}
where $\tilde{\gamma}^{1} = \pi^2 \chi^4/8$. 
Similarly the ratio $R^{(2)}$ of the second-loop contributions is 
\begin{equation}
\label{R2}
 R^{(2)} = \frac{L^{(2)}}
{\tilde{\gamma}^{1} \ln^4(s/M^2)} ~,
\end{equation}
where $\tilde{\gamma}^{2} = \pi^2 \chi^6/64$. 
Eqs.~(\ref{R1}, \ref{R2}) show explicitly that the difference 
between the one cut-off amplitude $\tilde{L}$ and the two cut-off 
amplitude $L_F^{(\mu)}$ grows with the order of the 
perturbative expansion, though rapidly decreasing with $s$. 

\begin{figure}[htbp]
 \begin{center}
 \includegraphics[width=9cm]{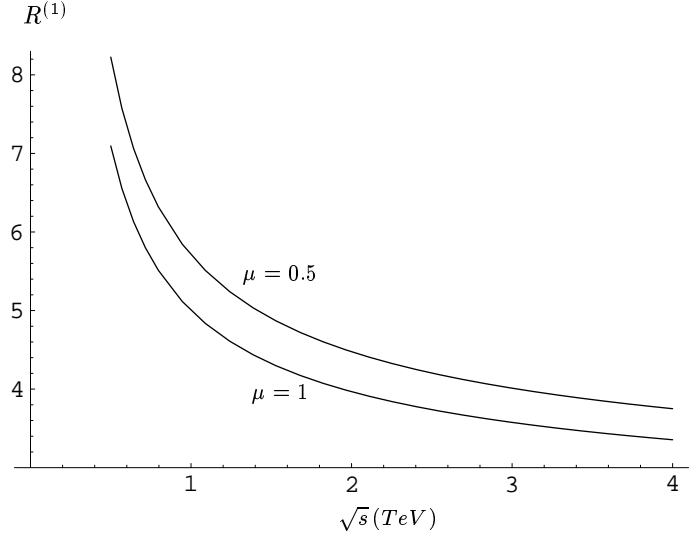}
 \caption{Dependence of $R^{(1)}$ on $s$ for different values of 
$\mu$(GeV).}
 \end{center}
\end{figure}
\begin{figure}[htbp]
 \begin{center}
 \includegraphics[width=9cm]{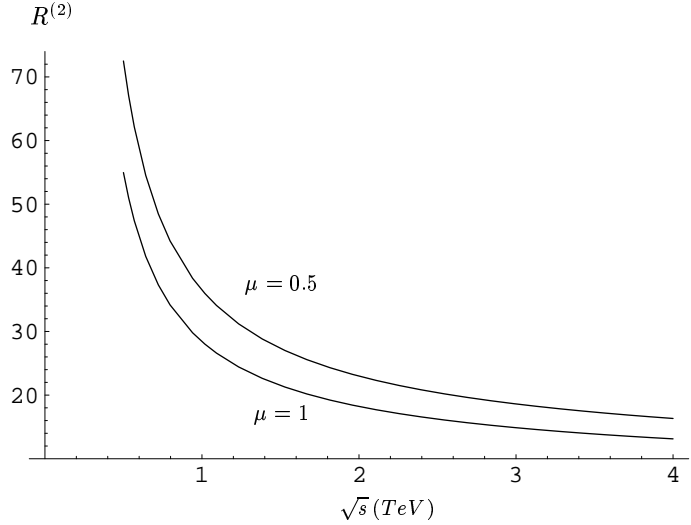}
 \caption{Dependence of $R^{(2)}$ on $s$ for different values of 
$\mu$(GeV).}
 \end{center}
\end{figure}
We can expect therefore that a sizable 
difference between $L_F^{(\mu)}$ and $\tilde{L}$ when all 
orders of the perturbative series are resumed. 
 
\section{Asymptotics of the forward scattering amplitude for $e^+e^-$ 
annihilation into $\mu^+ \mu^-$.}

In order to estimate the effect of higher order DL contributions on 
the difference between the one-cut-off and 
two-cut-off amplitudes, it is convenient to compare their 
high-energy asymptotics. For the sake of simplicity, we present below 
such asymptotical estimates for 
the amplitude $L_F^{\mu}$ of the forward $e^+e^-$ annihilation into 
$\mu^+ \mu^-$ in the collinear kinematics (\ref{tmu}). 
Calculations for the other 
amplitudes~(\ref{solutionaj}) can be done in a similar way. 
As well-known, the leading contribution to the asymptotic behavior is 
$L_F^{\mu} \sim s^{\omega_0}$, with ${\omega_0}$ being the 
rightmost singularity of the 
amplitude $L_F^{\mu}$.
This amplitude contains 
the amplitudes 
$\phi_{1,2}^{(0)}$ and $\phi_{1,2}$ 
and therefore also their singularities. 
Eqs.~(\ref{phi1}, \ref{phi0}) show that the 
singularities of both $\phi_1$ and $\phi_1^{(0)}$ are the square root 
branching points. 
The rightmost singularity of $\phi_1^{(0)}$ is $\chi_0$ 
and the rightmost singularity of $\phi_1$ is $\chi$. They are defined in 
Eqs.~(\ref{chi0}, \ref{chi}). 
Obviously, 
\begin{equation}
\label{phi01as}
\phi_1^{(0)}(\chi_0) = 4 \pi^2 \chi_0, 
~\phi_1^{(0)}(\chi) = 4 \pi^2 \big( \chi - \sqrt{\chi^2 - \chi_0^2}\Big)
 \equiv 4 \pi^2 \big( \chi - \chi'\big) , 
~\phi_1(\chi) = 4 \pi^2 \chi ~.
\end{equation} 

Combining Eqs.~(\ref{amplmu}) and (\ref{phi01as}) and neglecting the last 
integral in Eq.~(\ref{amplmu}), we obtain the 
asymptotic formula for the forward leptonic invariant amplitude $A$: 
\begin{equation}
\label{leptas}
L_F^{\mu} \sim 4 \pi^2 \Big(\frac{s}{\mu^2}\Big)^{\chi_0} \chi_0 + 4 \pi^2 
\Big(\frac{s}{M^2}\Big)^{\chi} 
\frac{2(\chi - \chi')(2\chi' - \chi) e^{2 \eta(\chi - \chi')}}
{3\chi - 2\chi' -(2\chi' - \chi) e^{2 \eta(\chi - \chi')}} .
\end{equation}

 The first term in Eq.~(\ref{leptas}) represents the 
asymptotic contribution of the QED Feynman graphs, the second term 
the mixing of QED and weak DL contributions. On the other hand, 
when the one-cut-off approach is used, 
the new amplitude ${\tilde{L}}_F^{\mu}$ asymptotically behaves as: 
\begin{equation}
\label{lepttilde}
 \tilde{L}_F^{\mu} \sim 4 \pi^2 \frac{\chi}{2}\Big(\frac{s}{M^2}\Big)^{\chi} ~.
\end{equation}
Then defining $Z(s, \eta)$, as:
\begin{equation}
\label{Z}
L_F^{\mu} = \tilde{L}_F^{\mu} \big( 1 + Z(s, \eta)\big) ~,
\end{equation} 
it is easy to see that 
\begin{equation}
\label{Zas}
Z (s, \eta) \sim \Big(\frac{s}{M^2}\Big)^{-\chi + \chi_0} 
\frac{2 \chi_0}{\chi} e^{\eta \chi_0} -1 +
\frac{4(\chi - \chi')(2\chi' - \chi) e^{2 \eta(\chi - \chi')}}
{\chi[3\chi - 2\chi' -(2\chi' - \chi) e^{2 \eta(\chi - \chi')}]} ~. 
\end{equation}
 
As $\chi_0 < \chi$, $Z(s)$ falls when $s$ grows. 
So, the one-cut-off and the two-cut-off approach lead to 
the same asymptotics, although at very high energies, say 
$\sqrt{s} \geq 10^6$~TeV. At lower energies, accounting for 
$Z$, the amplitude $L_F^{(\mu)}$ is increased by a factor of order 2. 
On the other hand, 
$Z$ strongly depends on the ratio $M/\mu$, which, of course, is related 
to the actual phenomenological conditions. To 
illustrate this dependence, we take $M = 100$~GeV and choose 
different values for $\mu$, ranging from $0.1$ to $1$~GeV. Then in 
Fig.~6 we plot 
$Z(s, \mu)$ 
for $\mu = 1$~GeV and $\mu = 0.5$~GeV. This shows that the variation is
approximately 1.5 at energies in the interval from 
$0.5$ to $5$~TeV. 
\begin{figure}[htbp]
 \begin{center}
 \includegraphics[width=9cm]{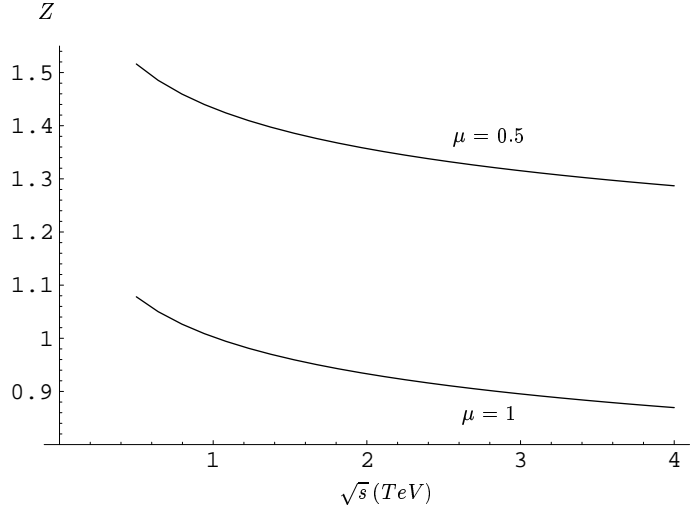}
 \caption{Dependence of $Z$ on $s$ for different values of $\mu$~(GeV).}
 \end{center}
\end{figure}

It is also interesting to estimate the difference between the purely 
QED asymptotics of $L_F^{\mu}$ 
(the first term in the rhs of Eq.~(\ref{leptas})) 
and the full electroweak asymptotics. To this aim, we introduce 
$\Delta_{EW}$: 
\begin{equation}
\label{deltaew}
L_F^{(\mu)} = \big(L_F^{(\mu)}\big)^{(QED)} (1 + \Delta_{EW})~. 
\end{equation} 

 From Eq.~(\ref{leptas}) we immediately get the following asymptotic 
behavior for $\Delta_{EW}$:
\begin{equation}
\label{deltaewas}
 \Delta_{EW} \sim \Big(\frac{s}{M^2}\Big)^{\chi - \chi_0} 
~~\frac{2(\chi - \chi')(2\chi' - \chi) e^{2 \eta(\chi - \chi')}}
{3\chi - 2\chi' -(2\chi' - \chi) e^{2 \eta(\chi - \chi')}} 
\end{equation} 
As $\chi > \chi_0$, $\Delta_{EW}$ grows with $s$, as shown 
in Fig.~7, Therefore 
the weak interactions contribution is 
approximately of the same size of the 
QED contribution, and 
 their ratio rapidly increases as $\mu$ decreases. 

\begin{figure}[htbp]
 \begin{center}
 \includegraphics[width=9cm]{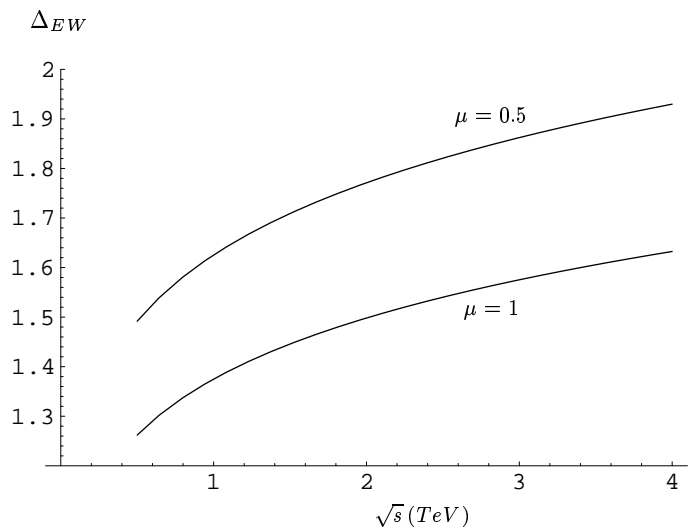}
 \caption{Dependence of $\Delta_{EW}$ on $s$ for different values of 
$\mu$~(GeV).}
 \end{center}
\end{figure}

\section{Summary and Outlook}
\label{CONCLUSIONS}

Next future linear $e^+e^-$ colliders will be operating in a energy
domain which is much higher than the electroweak bosons masses, so that
the full knowledge of the scattering amplitudes for $e^+e^-$
annihilation into fermion pairs will be needed.
In the present paper we have considered the high-energy non-radiative 
scattering amplitudes for $e^+e^-$ annihilation into leptons and quarks 
in the Regge kinematics~(\ref{tkin}) and (\ref{ukin}). 
We have calculated these amplitudes in the DLA, using a cut-off $M$, with 
$M \geq M_Z \approx M_W$, for the 
transverse momenta of virtual weak bosons and an
infrared cut-off $\mu$ for regulating DL contributions of virtual 
soft photons. We have obtained explicit 
expressions~(\ref{Phi1}, \ref{solutionaj}) for these amplitudes 
in the collinear kinematics~(\ref{tmu}, \ref{umu}) and 
Eqs.~(\ref{akappa}, \ref{akappam}) for the configuration where all 
Mandelstam variables are large. 
The basic structure of the expressions in the limit of collinear 
kinematics is quite 
clear. They consist of two terms: the 
first term 
presents the purely QED contribution, i.e. the one 
with virtual photon exchanges 
only, whereas the next term describe 
the combined effect of all electroweak boson exchanges. 
Obviously, 
in the limit when the cut-off $\mu \to M$, our 
expressions for the scattering amplitude converge to the much simpler 
expressions obtained in Ref~\cite{egt} with one universal 
cut-off for all electroweak bosons.
In order to calculate the electroweak scattering amplitudes, we 
derived and solved infrared equations for the evolution 
of the amplitudes with respect to the cut-offs 
$M$ and $\mu$. 
 
In order to illustrate the difference between the two methods, 
we have considered in more detail 
the scattering amplitude $L_F^{(\mu)}$ of the 
forward $e^+e^-$ annihilation into 
$\mu^+\mu^-$ and studied the ratios of the results obtained in the two 
approaches, first in one- and two-loop approximation and then to all 
orders to DLA. The ratios of the first- and 
second-loop DL results 
are plotted in Figs.~4 and 5. 
The total effect of higher-loop contributions is estimated 
comparing the asymptotic behaviors of the 
amplitudes. This is shown in Fig.~6. 
The effect of all electroweak 
DL corrections compared the QED ones is plotted 
in Fig.~7. It follows that accounting for all 
electroweak radiative corrections $L_F^{(\mu)}$ increases by up to 
factor of 2.5 at $\sqrt{s}\leq 1$~TeV, depending on the value of
$M/\mu$. 
In formulae for the $2 \to 2$ - electroweak cross sections, 
one can put $M = M_W \approx M_Z$ whereas the value 
of $\mu$ is quite arbitrary. However it vanishes, when these expressions 
are combined with cross sections of the radiative 
$2 \to 2 + X$ processes.

In the present paper we have considered the 
most complicated case of both the initial electron and 
the final quark or lepton being heft-handed (and their antiparticles 
right-handed). Studying other combinations of the 
helicities of the initial and final 
particles can be done quite similarly. 
We intend to use the results obtained in the present paper for further 
studying the forward-backward asymmetry at TeV 
energies, by including also the real radiative contributions. 
Basically, the QCD radiative corrections can give a big impact 
on the amplitudes of $e^+e^-$ - annihilation into hadrons. However, the 
perturbative QCD corrections cancel out of the expressions for the 
forward-backward asymmetry (see Ref.~\cite{egt}) whereas the non-perturbative corrections 
describing hadronization of the produced $q \bar{q}$ - pairs can 
be accounted for in the same way as it was done in Ref.~\cite{egt}. 

\section{Acknowledgement}
This work is supported by grants POCTI/FNU/49523/2002, 
SFRH/BD/6455/2001 and RSGSS-1124.2003.2


\begin{thebibliography}{99}

\bibitem{egt} B.I.~Ermolaev, M.~Greco and S.I.~Troyan. 
Phys.Rev. D 67(2003)014017.

\bibitem{flmm} V.S. Fadin, L.N. Lipatov, A.D.~Martin, and M.~Melles,
Phys.Rev. D 61(2000)094002.

\bibitem{pciaf} P.~Ciafaloni and D.~Comelli. 
Phys Lett B 476(2000)49.

\bibitem{me} M.~Melles Phys. Lett. B 495(2000)81 .
 
\bibitem{kod} M.~Hori, H.~Kawamura and J.~Kodaira. Phys. Lett. B 491(2000)275. 

\bibitem{bw} W.~Beenakker and A.~Werthenbach. Nucl. Phys. B 630(2002)3. 
 
\bibitem{dp} 
A.~Denner, M.~Melles and S.~Pozzorini. Nucl.Phys. B 662(2003)299.

\bibitem{kp} J.H.~Kuhn, A.A.~Penin and V.A.~Smirnov. 
Nucl.Phys.Proc.Suppl. 89(2000)94; Eur. Phys. J. C 17(2000)97.

\bibitem{m} S.~Moch. Nucl.Phys.Proc.Suppl. 116:23-27,2003.

\bibitem{ggfl} V.N.~Gribov, V.G.~Gorshkov, G.V.~Frolov, L.N.~Lipatov.
Sov.J.Nucl.Phys. 6(1968)95; ibid 6(1968)262.

\bibitem{Greco:1980mh}
M.~Greco, G.~Pancheri-Srivastava and Y.~Srivastava,
Nucl.\ Phys.\ B  171(1980)118
[Erratum-ibid.\ B 197(1982)543].

\end{thebibliography}
\end{document}